\documentclass{article}

\usepackage{color}

\begin{document}

\title{A correction on Shiloach's algorithm for minimum linear arrangement of trees}

\author{J. L. Esteban$^1$ and R. Ferrer-i-Cancho$^2$ \\
~\\
\begin{minipage}[c]{\textwidth}
{\small 
$^1$ Logic and Programming, LOGPROG Research Group, 
Departament de Ci\`encies de la Computaci\'o, Universitat Polit\`ecnica de Catalunya, Campus Nord, Edifici Omega. Jordi Girona Salgado 1-3. 08034 Barcelona, Catalonia, Spain \\
$^2$ Complexity \& Qualitative Linguistics Lab, LARCA Research Group,
Departament de Ci\`encies de la Computaci\'o, Universitat Polit\`ecnica de Catalunya, Campus Nord, Edifici Omega. Jordi Girona Salgado 1-3. 08034 Barcelona, Catalonia, Spain \\ 
}
\end{minipage}
}

\date{}

\maketitle

\begin{abstract}
More than 30 years ago, Shiloach published an algorithm to solve the minimum linear arrangement problem for undirected trees. Here we fix a small error in the original version of the algorithm and discuss its effect on subsequent literature. We also improve some aspects of the notation.
\end{abstract}



\section{Introduction}

Suppose that the $n$ vertices of a graph are sorted in a linear sequence and that the length of an edge is defined as the linear distance in the sequence between the vertices involved (adjacent vertices are at distance 1, vertices separated by a vertex are at distance 2 and so on). The minimum linear arrangement (m.l.a.) problem consists of finding the minimum sum of edge lengths over all the $n!$ sequences that can be formed \cite{Diaz2002}.
More formally, a linear arrangement $\pi$ is 1-to-1 a mapping of vertices onto $[1, n]$ such that $\pi(v)$ is the position of vertex $v$ in the sequence ($1 \leq \pi(v) \leq n$). Let $u \sim v$ indicate an edge between vertices $u$ and $v$. Then the sum of all edge lengths can be defined as
\begin{equation}
D = \sum_{u\sim v} |\pi(u) - \pi(v)|,
\end{equation}  
where $|\pi(u) - \pi(v)|$ is the length of $u\sim v$.
Solving the m.l.a. problem consist of finding $D_{min}$, the minimum $D$ over all the possible linear arrangements. Although the solution of the m.l.a. problem is an NP-hard optimization problem in general, polynomial time algorithms for undirected trees do exist \cite{Shiloach1979,Chung1984}.

More than 30 years ago, Shiloach published an $\mathcal{O}(n^{2.2})$ algorithm to solve the m.l.a. problem for undirected trees \cite{Shiloach1979}. 
A few years later, 
Chung published two different algorithms for solving the same problem \cite{Chung1984}. The first one has cost $\mathcal{O}(n^2)$ and it is quite similar to Shiloach's algorithm. The second one has cost $\mathcal{O}(n^\lambda)$, where $\lambda>\log 3/\log2$. 
To our knowledge, Chung's second algorithm is still the most efficient algorithm for undirected trees. This is corroborated by surveys \cite{Petit2011a, Lai1999a, Diaz2002}. 
As far as we know, these algorithms have not been implemented and tested. We implemented Shiloach's algorithm and found an error, which is the subject of this note.

We discovered this mistake trying to understand why our implementation of Shiloach's algorithm \cite{Esteban2016a} failed for complete binary trees of $k$ levels with $k \geq 5$. For trees with $k \geq 1$, the solution of the m.l.a. is \cite{Chung1978a}
\begin{equation}
D_{min} = 2^k \left(\frac{k}{3} + \frac{5}{18}\right) + (-1)^k \frac{2}{9} - 2.
\label{complete_binary_trees_equation}
\end{equation}
For $k=5$, Eq. \ref{complete_binary_trees_equation} gives $D_{min} = 60$ while our original implementation of Shiloach's algorithm gave $D_{min} = 46$. 
Once we corrected the mistake, our implementation of Shiloach's algorithm ceased to give wrong results.

The remainder of the article is organized as follows. Section \ref{error_section} presents the error and two possible corrections. Section \ref{notation_section} improves some aspects of the notation of Shiloach's article. Section \ref{discussion_section} discusses some implications of the correction for related work.

\section{The error and its correction}
\label{error_section}

Let $T$ be an undirected tree with $n$ nodes, and $v_*$ one of its nodes. Let $T_0,T_1,\dots,T_k$ be the subtrees of $T$ that are produced when $v_*$ and its edges are removed from $T$. 
Let $n_i$ be the number of nodes of $T_i$. $T_0,T_1,\dots,T_k$ are sorted by decreasing size, i.e. $n_0\ge n_1\ge\dots\ge n_k$. $v_0,v_1,\dots,v_k$ are the vertices of each subtree that are connected to $v_*$ in $T$. 
A type $B$ arrangement of a tree $T(\alpha)$, as defined in Theorem 3.1.1 b), depends on a certain calculated parameter $p_\alpha$ (where $\alpha$ is either 0 or 1), and consists of placing the tree $T_* = T(\alpha) - (T_1,...,T_{2p_\alpha - \alpha})$ at the center surrounded by subtrees $T_1,...,T_i, ..., T_{2p_\alpha - \alpha}$ as indicated in Figs. 4(a) and 4(b) of Shiloach's article.
$\alpha = 0$ indicates that the tree $T$ is a free tree (Fig. 4(a)) and $\alpha = 1$ indicates that $T$ is an anchored tree (Fig. 4(b)).
Recall section 2.2 and equations (2.2) and (2.3) from \cite{Shiloach1979} for the definition and costs of left and right anchored trees. 

Section 3.1. (p. 18) defines $n_*$ as 
\begin{equation}
n_* = n - \sum_{i=0}^{2p_\alpha - \alpha} n_i,
\label{size_of_central_tree_equation}
\end{equation} 
The error has to do with the use of $n_*$ in the calculation of the cost of an arrangement of type B.
Part b) of Theorem 3.1.2 defines the cost of an arrangement of this type for a tree $T$ as\footnote{We have copied the equation from \cite{Shiloach1979}. Besides the error we are correcting, there is a mistake in the use of $\pi_i$ that will be addressed in Section \ref{notation_section}}  
\begin{eqnarray}
C_\alpha(B) = C[\pi, T(\alpha)] & = & \sum_{\footnotesize\begin{array}{c} i=1 \\ i \mbox{~is odd} \end{array}}^{2p_\alpha - 1} C[\pi, \overrightarrow{T_i}(v_i)] +  \sum_{\footnotesize \begin{array}{c} i=1 \\ i \mbox{~is even} \end{array}}^{2p_\alpha - 2\alpha} C[\pi_i, \overleftarrow{T_i}(v_i)] + \nonumber \\ 
                           &   & C[\pi, T_*] + S_\alpha.\label{type_B_arrangement_equation}
\end{eqnarray}  
$\pi$ is an arrangement for the whole $T(\alpha)$ that 
it is constructed in such a way that the nodes of any subtree $T_i$ are mapped into $n_i$ consecutive numbers.
Note that all the summands of the form $C[\pi,\overrightarrow{T_i}(v_i)]$ measure the cost of $\overrightarrow{T_i}(v_i)$ as an independent tree 
with $n_i$ nodes and a right anchor; all the summands in the form $C[\pi,\overleftarrow{T_i}(v_i)]$ measure the cost of $\overleftarrow{T_i}(v_i)$ as an independent tree with $n_i$ nodes and a left anchor.
In order to account for the cost of the whole $T$ we need also to take into account $S_\alpha$, i.e. the cost of joining all the anchored subtrees with $T_*$. To achieve this goal $S_\alpha$ is added in Eq. \ref{type_B_arrangement_equation}.

At the bottom of p. 18, Shiloach defines  
\begin{eqnarray*}
S_0 = ... + p_0(n_* + 1) \\
S_1 = ... + p_1(n_* + 1) - 1.
\end{eqnarray*}  
If we follow the definition of $n_*$ in \cite{Shiloach1979} (Eq. \ref{size_of_central_tree_equation}), an accurate derivation of $S_0$ and $S_1$ (see below) indicates that their definitions should read as 
\begin{eqnarray*}
S_0 = ... + p_0(n_* + n_0 + 1) \\
S_1 = ... + p_1(n_* + n_0 + 1) - 1.
\end{eqnarray*}
This little mistake implies that part b) of Theorem 3.1.2 is wrong. The error concerns steps 6-7 of Shiloach's algorithm (Section 3.2 of his article, pp. 19-20). 
Shiloach omitted the proof of that theorem arguing that it "is by a straightforward  calculation which follows from elementary definitions" (pp. 19). 

Let us consider the subtree $\overrightarrow{T_i}(v_i)$. In this case, $C[\pi, \overrightarrow{T_i}(v_i)]$ includes the cost (or length) of the anchor of $\overrightarrow{T_i}(v_i)$, but notice that the cost of the anchor is only a part of the cost of the edge joining node $v_i$ to $T_*$. Thus, the couple of summations in Eq. \ref{type_B_arrangement_equation} comprise the cost of the anchors of every anchored subtree $T_i$, but the edge joining any $T_i$ to $T_*$ is longer that the anchor of $T_i$. $S_\alpha$ is added to account for the missing part of the cost, which in case that $\alpha = 1$ also has to account for the cost of the anchor of $T$. 
When $\alpha = 0$, $T_*$ has $p_0$ subtrees to its left and $p_0$ subtrees to its right. In contrast, when $\alpha = 1$, $T_*$ has $p_1$ subtrees to its left and $p_1-1$ subtrees to its right. With this background in mind, a derivation of $S_0$ and $S_1$ is straightforward.
 
With the help of Fig. 4(a) and the definition of $C_\alpha(B)$, one obtains
\begin{displaymath}
S_0 = (n_3 + n_4) + 2(n_5 + n_6) + (p_0 - 1)(n_{2p_0 - 1} + n_{2p_0}) + p_0(Z + 1), 
\end{displaymath}
where $Z$ is the number of vertices of the tree $T_*$. If $Z = n_*$ we get exactly Shiloach's definition of $S_0$. 
The problem is that $T_0$ is a subtree of $T_*$, but recalling the definition of $n_*$ in Eq. \ref{size_of_central_tree_equation} we see
that the size of $n_0$ is substracted, so $Z = n_* + n_0$. 
Recalling that $S_1$ comprises the length of the anchor of $T(\alpha)$, Fig. 4(b) helps one to see that  
\begin{displaymath}
S_1 = (n_2 + n_3) + 2(n_4 + n_5) + (p_1 - 1)(n_{2p_1 - 2} + n_{2p_1 - 1}) + p_1(Z + 1) - 1,
\end{displaymath}
where $Z = n_* + n_0$ again. 

Above we have explained the error and fixed it changing $S_0$ and $S_1$. Alternatively, the error can be fixed redefining $n_*$ as 
\begin{displaymath}
n_* = n-\sum_{i=1}^{2p_\alpha-\alpha}n_i.
\end{displaymath}
Notice that we do not substract $n_0$ from $n$ as in the original definition of $n_*$ in Eq. \ref{size_of_central_tree_equation}. This new $n_*$ is exactly the number of nodes of $T_*$, thus it makes right the original definition of $S_0$ and $S_1$. Interestingly, the new definition of $n_*$ is a natural extension of the definition of $n_i$ as the size of $T_i$ for $0\le i \le k$ (main text) or $n_\alpha$ as the size of $T_\alpha$ (footnote) in p. 16 of Shiloach's article.

As we are changing the original definition of $n_*$, a further correction is necessary in the definition of $p_\alpha$ in Section 3.1 (p. 18) of \cite{Shiloach1979}. In particular, 
\begin{displaymath}
n_i > \left\lfloor\frac{n_0+2}{2}\right\rfloor + \left\lfloor\frac{n_*+2}{2}\right\rfloor
\end{displaymath}
must be replaced by  
\begin{displaymath}
n_i > \left\lfloor\frac{n_0+2}{2}\right\rfloor + \left\lfloor\frac{n_*-n_0+2}{2}\right\rfloor
\end{displaymath}
in order to calculate $p_\alpha$ correctly. 
In the next section, we will assume this redefinition of $n_*$ that is involved in the definition of $p_\alpha$.




\section{A question of notation}
\label{notation_section}

Eq. \ref{type_B_arrangement_equation} is copied as is from \cite{Shiloach1979}. There is a problem with notation in this equation.
Note that the terms in the first summation are $C[\pi, \overrightarrow{T_i}(v_i)]$ while the terms in the second summation are like $C[\pi_i, \overleftarrow{T_i}(v_i)]$, namely the former summation uses the arrangement $\pi$ while the other uses $\pi_i$ as if it were a modification of the arrangement $\pi$ tailored to a subtree, but $\pi_i$ is never defined. The inconsistent use of $\pi$ and $\pi_i$ in both summations is clearly a typo. 

We introduce $[a,b]=\{a,a+1,\dots,b-1,b\}$. Note that $T(\alpha)$ is a tree with $n$ nodes, and $\pi$ is an arrangement for $T(\alpha)$ and thus $\pi$ is a bijective mapping from $[1,n]$ to $[1,n]$.
We can use $\pi$ to calculate $C[\pi, T_*]$ because $T_*$ is not anchored. For the remainder of subtrees $T_i$ in Eq. \ref{type_B_arrangement_equation}, $\pi$
should be used with caution to calculate the cost because these trees are anchored. In general, the cost of a right anchored tree $\overrightarrow{T}(v)$ (Eq. 2.2 of Shiloach's article) is  
\begin{equation}
C[\pi, \overrightarrow{T}(v)] = C[\pi, T] + n - \pi(v),
\label{cost_of_right_anchored_tree}
\end{equation}
where $n - \pi(v)$ is the length of the anchor and $n$ is the size of $T$.  
Coming back to $T_i$, the length of the anchor depends on $n_i$ (not on $n$), and also depends on $\pi(v_i)$, but note that the value of $\pi(v_i)$ takes into account the size of all the trees to the left of $T_i$ in the arrangement and $\pi(v_i) \notin [1, n_i]$ except for $T_1$, that has no tree before it in the arrangement. That means that $\pi(v_i)>n_i$ and thus $\pi(v_i)$ cannot be used to calculate the length of the anchor with Eq. \ref{cost_of_right_anchored_tree}, except for $T_1$. Similar problems apply to left anchored trees. 
Recalling the definition of a type $B$ arrangement (Fig. 4 of Shiloach's article), the sum of the sizes of the trees to the left of $T_i$ is 
\begin{displaymath}
left(i) = 
   \left\{
      \begin{array}{ll}
         \sum_{j=1, j\ odd}^{i-1}n_j & \mbox{if $i$ is odd} \\ 
         n -\sum_{j=2,j\ even}^{i}n_j  & \mbox{if $i$ is even}.
      \end{array}
   \right.
\end{displaymath}
for any $T_i$ to the left or to the right of $T_*$.
A possible way of unifying the notation is defining $\pi_a$ as a bijective mapping from $[1,n]$ to $[1-a,n-a]$, where 
\begin{equation}
\pi_a(i) = \pi(i) - a.
\label{new_linear_arrangement_equation}
\end{equation}
For any $a$, $C(\pi,T)=C(\pi_a,T)$. 
For right anchored trees, Eqs. \ref{cost_of_right_anchored_tree} and \ref{new_linear_arrangement_equation} give
\begin{eqnarray*}
C[\pi,\overrightarrow{T}(v)] & = & C[\pi_a,T] + n - \pi_a(v) -a \\
                             & = & C[\pi_a,\overrightarrow{T}(v)]-a 
\end{eqnarray*}  
and similarly for left anchored trees. Recalling Fig. 4 of Shiloach's article, we apply the new arrangement to rewrite Eq. \ref{type_B_arrangement_equation} as 
 \begin{eqnarray}
C_\alpha(B) & = & C[\pi, T(\alpha)] \nonumber \\
            & = & \sum_{\footnotesize \begin{array}{c} i=1 \\ i \mbox{~is odd} \end{array}}^{2p_\alpha - 1} C[\pi_{left(i)}, \overrightarrow{T_i}(v_i)] + \nonumber \\ 
            &   & \sum_{\footnotesize \begin{array}{c} i=1 \\ i \mbox{~is even} \end{array}}^{2p_\alpha - 2\alpha} C[\pi_{left(i)}, \overleftarrow{T_i}(v_i)] + \nonumber \\ 
            &   & C[\pi, T_*] + S_\alpha.\label{type_B_arrangement_equation_new}
\end{eqnarray}  
Notice that $\pi_{left(i)}$ is mapping the nodes of $T_i$ to $[1,n_i]$ and thus the length of the anchor is calculated correctly with $n = n_i$ and $\pi(v) = \pi_{left(i)}(v_i)$ for any $T_i$ according to Eqs. 2.2 and 2.3 of Shiloach's article.

The separate but similar definitions of $S_0$ and $S_1$, can be expressed compactly and equivalently with a single definition of $S_\alpha$, i.e. 
\begin{eqnarray*}
S_\alpha & = & (n_{3 - \alpha} + n_{4 - \alpha}) + 2(n_{5 - \alpha} + n_{6 - \alpha}) + \dots \\ 
         &   & + (p_\alpha - 1)(n_{2p_\alpha - 1 - \alpha} + n_{2p_\alpha - \alpha}) + p_\alpha(n_* + 1) - \alpha.
\end{eqnarray*}

The improved notation for arrangements has implications for Theorem 3.1.2 a), that originally defined 
\begin{displaymath}
C_\alpha(A) = C[\pi,T(\alpha)] = \left\{ 
                                    \begin{array}{ll}
                                    C[\pi,\overrightarrow{T_0}(v_0)] + C[\pi, \overleftarrow{T-T_0}(v_*)] + 1. & \mbox{if~}\alpha = 0 \\
                                    C[\pi,\overrightarrow{T_0}(v_0)] + C[\pi, T - T_0] + n - n_0 & \mbox{if~}\alpha = 1.
                                    \end{array}
                                 \right. 
\end{displaymath}
The only necessary change concerns the case $\alpha = 0$. In particular, the second $C[\pi,...]$ has to be replaced by $C[\pi_{n_0},...]$. The explanation is as follows (recall Fig. 4 of Shiloach's article). First let us consider the case $\alpha=0$. For $C[\pi,\overrightarrow{T_0}(v_0)]$, $\pi$ maps the $n_0$ nodes of $T_0$ to $[1,n_0]$ and thus it is not necessary to change $\pi$. 
For $C[\pi, \overleftarrow{T-T_0}(v_*)]$, $\pi$ has to be replaced by $\pi_{n_0}$ because the vertices of $T-T_0$ are placed after the vertices of $T_0$. Therefore 
\begin{displaymath}
C_\alpha(A)=C[\pi,T(\alpha)]=C[\pi,\overrightarrow{T_0}(v_0)] + C[\pi_{n_0},\overleftarrow{T-T_0}(v_*)]+1.
\end{displaymath} 
When $\alpha=1$, no change is needed because $\pi$ is suitable for $T_0$ as before and also for $T-T_0$ because this tree is not anchored. 

\section{Discussion}
\label{discussion_section}

We have seen two ways of correcting Shiloach's algorithm: one that is based on the original definition of $n_*$ and another based on a more elegant definition. Related research must be revised in light of the small error that we have reported above. 

Chung's first algorithm is similar to Shiloach's: for certain values $p$ and $q$, which play a role similar to Shiloach's $p_\alpha$, Chung's first algorithm arranges vertices placing the tree $T_* = T - (T_{i_1},...,T_{i_{2p+1}})$ at the center surrounded by subtrees $T_{i_1},..., T_{i_{2p+1}}$, or placing the
tree $T_* = T - (T_{i_1},...,T_{i_{2q}})$ at the center surrounded by subtrees $T_{i_1},..., T_{i_{2q}}$. 
In Chung's first algorithm, calculations that are equivalent to Shiloach's $S_0$ and $S_1$ 
appear within Properties 12 and 13 (p. 46). 
In particular, the bit 
\begin{displaymath}
ns-\sum_{j=1}^{s} (s-j+1)(t_{i_j}+t_{i_{2s-j+1}})
\end{displaymath}
in Property 12 corresponds to $S_0$. 
The bit 
\begin{displaymath}
n(s+1)-\sum_{j=1}^{s} (s-j+1)(t_{i_j}+t_{i_{2s-j+1}}) - (s+1)t_{i_{2s+1}}
\end{displaymath}
in Property 13 corresponds to $S_1$. Properties 12 and 13 are used, respectively, in Step 4 and Step 7 of Chung's first algorithm. While Shiloach calculates $S_0$ and $S_1$ by summation and omits one number in each, Chung proceeds by substraction from a maximum and omits no number. Furthermore, we have checked both properties and we find them correct. Chung's second algorithm (the one with subquadratic cost) also uses Properties 12 and 13, which are correct. Therefore, we conclude that Chung's algorithms are not affected by the error in Shiloach's algorithm. 

Beyond Shiloach's and Chung's algorithm, we expect that the error in Shiloach's algorithm does not affect or can be easily fixed because it concerns a very specific component of the algorithm.  
For instance, Shiloach's algorithm was parallelized by D\'iaz and colleagues \cite{Diaz1997a}. The error does not affect their conclusions. Just correcting the formulae as indicated before suffices.

\section*{Acknowledgements}
We are very grateful to L. J. Schulman and two anonymous reviewers, who have helped us to improve the article significantly. 
We also thank M. Serna and J. Petit for their advice. 
JLE is funded by the project TASSAT3 (TIN2016-76573-C2-1-P) from MINECO (Ministerio de Economia y Competitividad). 
RFC is funded by the grants 2014SGR 890 (MACDA) from AGAUR (Generalitat de Catalunya) and also
the APCOM project (TIN2014-57226-P) from MINECO. 

\bibliographystyle{unsrt}

\bibliography{cor}

\begin{thebibliography}{1}

\bibitem{Diaz2002}
J.~D\'iaz, J.~Petit, and M.~Serna.
\newblock A survey of graph layout problems.
\newblock {\em ACM Computing Surveys}, 34:313--356, 2002.

\bibitem{Shiloach1979}
Y.~Shiloach.
\newblock A minimum linear arrangement algorithm for undirected trees.
\newblock {\em SIAM J. Comput.}, 8(1):15--32, 1979.

\bibitem{Chung1984}
F.~R.~K. Chung.
\newblock On optimal linear arrangements of trees.
\newblock {\em Comp. \& Maths. with Appls.}, 10(1):43--60, 1984.

\bibitem{Petit2011a}
J.~Petit.
\newblock Addenda to the survey of layout problems.
\newblock {\em Bulletin of the European Association for Theoretical Computer
  Science}, 105:177--201, 2011.

\bibitem{Lai1999a}
Y.-L. Lai and K.~Williams.
\newblock A survey of solved problems and applications on bandwidth, edgesum,
  and profile of graphs.
\newblock {\em Journal of Graph Theory}, 31(2):75--94, 1999.

\bibitem{Esteban2016a}
J.~L. Esteban, R.~{Ferrer-i-Cancho}, and C.~G\'omez-Rodr\'iguez.
\newblock The scaling of the minimum sum of edge lengths in uniformly random
  trees.
\newblock {\em Journal of Statistical Mechanics}, page 063401, 2016.

\bibitem{Chung1978a}
F.~R.~K. Chung.
\newblock A conjectured minimum valuation tree.
\newblock {\em SIAM Review}, 20:601~604, 1978.

\bibitem{Diaz1997a}
J.~D\'iaz, A.~Gibbons, G.~E. Pantziou, M.~J. Serna, P.~G. Spirakis, and
  J.~Toran.
\newblock Parallel algorithms for the minimum cut and the minimum length tree
  layout problems.
\newblock {\em Theoretical Computer Science}, 181(2):267 -- 287, 1997.

\end{thebibliography}
  
\end{document}